\documentstyle[twoside,fleqn,espcrc2,epsfig]{article}

\newcommand{\AmS}{{\protect\the\textfont2
  A\kern-.1667em\lower.5ex\hbox{M}\kern-.125emS}}
\title{Diffractive photoproduction of $\Upsilon$ states at HERA}

\author{M. F. McDermott\address{Dept. of Physics and Astronomy, Schuster Lab., Brunswick St., University of Manchester, Manchester, England}%
        \thanks{In collaboration with M. Strikman and L. Frankfurt}}

\begin{document}

\begin{abstract}
Cross sections for the diffractive photoproduction of the 
$\Upsilon$-family at HERA energies, within the 
framework of the analysis by Frankfurt, K\"{o}pf
and Strikman \cite{FKS1,FKS2}, are presented.
They compare well with the recent  preliminary data from ZEUS and H1.
Two novel effects lead to a significant enhancement 
of the original calculation: the non-diagonal (or skewed) kinematics, 
calculated to leading-log($Q^2$) accuracy, 
and the large magnitude of the real part of the amplitude. 
A considerably stronger rise in energy is predicted than that found in 
$J/\psi$-production.
\end{abstract}

\maketitle

\section{Basic Formulae}

\begin{figure}[htb]
\vspace{9pt}
\begin{center}
\mbox{
\epsfxsize = 5.0cm
\epsfbox{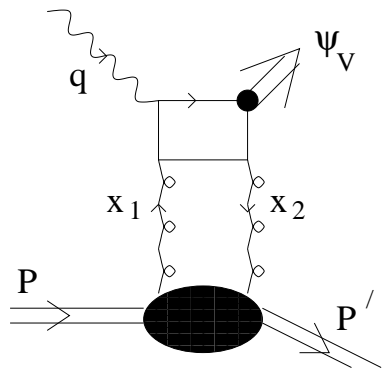}}
\end{center}
\caption{Photoproduction of heavy vector mesons}    
\label{figups}
\end{figure}

Diffractive heavy vector meson photo- and electro- production are governed by 
the exchange of two gluons, in a colour singlet, in the $t$-channel 
(see figure \ref{figups}). 
The amplitude for this hard diffractive process involves the overlap of the scattering 
cross-section for small dipoles (specified by momentum sharing, $z$, and transverse size, $b^2$) with the light-cone wavefunctions for the photon and vector meson:

\begin{equation}
{\cal A} \propto \int dz \int d^2 b \psi_{\gamma} (z,b) \hat{\sigma} (b^2) 
\psi_{V} (z,b).
\label{eq1}
\end{equation}

The universal cross section for the scattering a small dipole off the proton
is directly proportional to its transverse size, and the gluon density
\begin{equation}
 \hat{\sigma} (b^2) = \frac{\pi^2}{3} b^2 \alpha_{s} (b^2) x g (x,b^2)
\label{eq2}
\end{equation}

We exploit this universality of $\hat{\sigma}$
to set the relation between transverse dipole size and four-momentum scales
by expressing $\sigma_L$ as a similar integral involving 
$\hat{\sigma}$ convoluted with the square of the (known) longitudinally polarized 
photon wavefunction, $\psi^{L}_{\gamma} (z,b)$. For a given $x,Q^2$ we calculate the
average $b^2$ of this integral and, employing the ansatz $b^2 = \lambda/Q^2$,
determine the relationship by an iterative proceedure.
It turns out that $\lambda$  has a only weak dependence on $x,Q^2$ \cite{FMS},
which gives us faith in our ansatz. 

In the integral in equation 1, $\alpha_s xg$ is a slow function of $b^2$, 
so may be taken out of the integral at an average point.
The average $<\! \! b^2 \! \!>$ is taken to be the median of this integral 
as explained in \cite{FMS}, the effective four-momentum scale of a particular production
process is then $Q^2_{eff} = \lambda/\! <\! \! b^2 \! \!>$. Different wavefunctions weight the integrand differently, yielding a $Q^2_{eff}$ which reflects this.

The cross section for the photoproduction of heavy vector-meson states is then

\begin{eqnarray}
\sigma(\gamma P \rightarrow V P) & = &
\frac{3 \pi^3 \Gamma M_V^2 (1 + \beta^2)}{64 \alpha_{em} (m_q^2)^4 B_{D,V}} \times \nonumber \\
 C(Q^2=0) & \times  & \left[\alpha_s (Q^2_{eff}) g(x_1,\delta,Q^2_{eff})\right]^2 \label{eq3}
\end{eqnarray}

\noindent where $M_V, \Gamma, B_{D,V}$ are the mass, decay width to 
leptons and diffractive slope of the vector meson concerned. The factor $C(Q^2=0) < 1$
contains the remaining integral (squared) suitably normalised 
(see \cite{FKS1,FKS2,FMS} for more details). 
Information about the real part 
of the amplitude is contained in $\beta = {\cal R}e {\cal A} / {\cal I}m {\cal A}$. 
The fact that one must create a large time-like mass from a real 
photon ensures that the process of figure \ref{figups} is off-diagonal ($x_1 \neq x_2$),
as specified by the ``skewedness'' parameter $\delta = x_1 - x_2 = M_V^2/W^2$. This leads to the replacement of the ordinary gluon distribution by
its skewed generalization in equation \ref{eq2}.

\section{Key Issues}

The use of the naive QCD cross section in DLLA \cite{RYS} 
(static wavefunctions, no skewedness or rescaling, etc) 
disagrees with the first ZEUS \cite{ZEUS} and H1 \cite{H1} data 
by a factor of 5-10, and it is necessary to take corrections to this 
asymptotic formula into account. 

The experimental signal is {\it exclusive diffractive}: only the muons, to which 
the $\Upsilon$ states decay, are observed in the main detector. 
The resolution of the muon chambers is such that the three $S$-states are not resolved,
 so the actual measurement corresponds to the sum of the products of the 
production cross sections for each state times their respective branching ratios to muons. 

In calculating the cross section we use the hard (small average $b^2$) 
hybrid wavefunctions of \cite{FKS2} for the vector mesons. These 
are given by boosted Schr\"{o}dinger quarkonium wavefunctions, modified at 
small $b^2$ to impose QCD-type behaviour ($z(1-z)$) 
(and normalised to the decay width to leptons).
These hard wavefunctions weight the $b$-integral in equation \ref{eq1} 
towards smaller $b^2$ where it is suppressed by the $b^2$ factor in equation \ref{eq2}.
This leads to a $k_{T}^2$-suppression (encoded in $C$) which is tempered 
by the related larger value of $Q^2_{eff}$ at which the skewed gluon density is sampled. Typical values for the effective scale, at 
$x \approx 0.01$ are $Q^2_{eff} = 40, 62, 76$ GeV$^2$ for 
$\Upsilon, \Upsilon^{'}, \Upsilon^{''}$ respectively, and it has a 
fairly weak $x$-dependence \cite{FMS}. 

The skewed kinematics of the amplitude cause us to replace the ordinary gluon density in
equation \ref{eq3} by its skewed, or off-diagonal equivalent $g(x_1,\delta,Q^2_{eff})$. 
In practice one must make an ansatz for the skewed distribution at the  starting scale, 
and evolve using skewed splitting functions (known only to leading order at present). 
In practice, we follow \cite{fel} and assume that the dependence on $\delta$ may be neglected 
at the starting scale. However, for a sufficiently long evolution 
(as in this case, with large $Q^2_{eff}$) 
one becomes increasingly insensitive to the details of this starting assumption 
and the difference between skewed and conventional is driven by the evolution.
For the leading order partons that one must use for consistency at present,
this leads to an overall enhancement factor of about 
2-2.6 in the cross sections for $\Upsilon$-states.

The fact that in $\Upsilon$-photoproduction at HERA one is sampling the skewed gluon at 
rather large scales and relatively high $x$, led us to re-examine the calculation of 
$\beta$. Usually it is sufficient to relate it to the logarithmic derivative of the gluon 
distribution, which is valid if this quantity may be fitted by a simple (small) power in $x$ (this works for the $J/\psi$ case). We found that for $\Upsilon$-states,
 certainly for the lower values of $W^2$, it was more appropriate to 
use a two power fit to the skewed distribution and use dispersion relations 
to get to $\beta$. Overall the two-power fit leads to an additional factor of about 2. 
A numerical comparison between the two methods may be found in table 5 of \cite{FMS}.

In practice all of these effects are calculated precisely and systematically, 
at a given value of $W^2$, within the computer codes 
to produce the final result.

\section{Results and discussion}

\begin{figure}[htb]
\vspace{9pt}
\mbox{
\epsfxsize = 7cm
\epsfbox{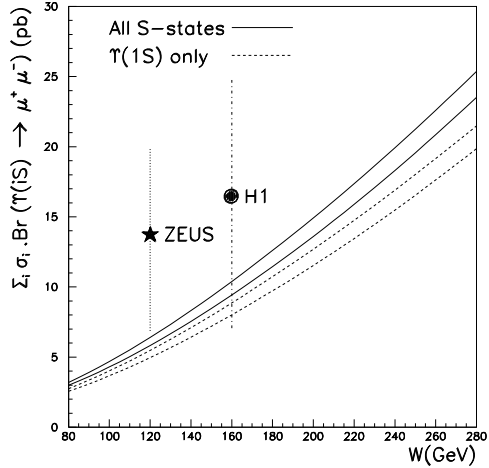}
}
\caption{Sum of cross sections times branching ratios to muons for $\Upsilon$(1s), $\Upsilon^{'}$(2s), $\Upsilon^{''}$(3s) as a function of energy. The dashed curves are for $\Upsilon $(1s) alone. The upper (lower) curves in each case correspond to MRSTLO (CTEQ4L) partons at the starting scale. Also shown are the data from ZEUS and H1 (preliminary), at their respective mean energies, with systematic and statistical errors added in quadrature.}
\label{figsum}
\end{figure}

Taking all the above corrections into account leads to QCD cross section which are in fair agreement with the first data: see figure \ref{figsum}. The following values have been  
calculated for the $C$-factors of equation \ref{eq3}: $C(Q^2=0) = 0.29, 0.14, 0.08$ for 
$\Upsilon, \Upsilon^{'}, \Upsilon^{''}$, respectively.
This implies that within our model $\Upsilon(1s)$ is responsible 
for about $85 \%$ of the signal, in contrast to the $70 \%$ 
assumed by the experiments in extracting their $\Upsilon$(1s) cross sections. 
As can be clearly seen from the figure a very strong rise with energy 
($ W^{1.6-1.7} $) is expected. The is being driven by the (square) of the
skewed gluon distribution, sampled at large effective scales, $Q^2_{eff}$.

Ryskin {\it et al} \cite{MRT} recently released a preprint on this topic which is in 
qualitative agreement with our results for the cross sections. 
However the details of the contributing correction factors of the 
previous section are different, with the main disagreement coming from the 
treatment of the wavefunction for the vector meson (see \cite{TT} for more details). 
It is hoped that a global analysis of all hard diffractive vector meson data
will help to pin down these uncertainties. Of particular importance is 
the explicit calculation of the contribution of the higher order Fock states 
in the vector meson wavefunctions. 
Ratios of cross section within families as a function of $Q^2$ and $W^2$, 
e.g. $\psi^{'}/J/\psi$, are expected to be particularly sensitive
to this issue (they should also reveal the effective scales involved).

While the presence of skewedness $\delta= (Q^2 + M_V^2)/(W^2+Q^2)$ 
is a purely kinematical effect, the enhancement it provides at small x 
and large momentum scales is due solely to QCD evolution.
As such, the enhancement is expected to be a common feature of  
other exclusive processes which share the same kinematics, for 
example the  electroproduction of $\rho, J/\psi$ at HERA.  
All of these issues are under active investigation at present \cite{FMS2}.
 
Given the quantum mechanical nature of effect of skewedness, it would be interesting to see if one can find a process which is sensitive to destructive interference effects as well as the constructive interference effects found here. 

With the increasingly precise data being released from H1 \cite{H12} and ZEUS \cite{ZEUS2}, heavy vector meson production promises to remain a fascinating 
area in the next few years.

\end{document}